# Study of activation cross sections of deuteron induced reactions on erbium in the 32-50 MeV energy range

F. Tárkányi[a], A. Hermanne[b], F. Ditrói[a*], S. Takács[a]

[a] Institute for Nuclear Research, Hungarian Academy of Sciences (ATOMKI), 4026 Debrecen, Hungary

[b] Cyclotron Laboratory, Vrije Universiteit Brussel (VUB), 1090 Brussels, Belgium

## Abstract

Activation cross sections of the $^{nat}Er(d,x)^{163,165,166,167,168,170}Tm$ and $^{nat}Er(d,x)^{171,161}Er$ nuclear reactions have been measured in the 32-50 MeV energy range, above 40 MeV for the first time. The activation method with stacked foil irradiation technique and gamma-ray spectroscopy were used. The experimental cross sections were compared with the theoretical predictions in the TENDL-2015 library.

* Corresponding author: ditroi@atomki.hu

# 1. Introduction

We perform systematic study of excitation functions of light ion induced nuclear reactions on rare earth isotopes, to investigate production routes of diagnostic and therapeutic radioisotopes with potential use in nuclear medicine.

In case of erbium target the activation cross sections of alpha particle induced reactions up to 40 MeV (Király et al., 2007, 2008) and the proton induced reactions up to 70 MeV (Hermanne et al., 2011a; Tarkanyi et al., 2016; Tárkányi et al., 2008, 2009) have already been reported by us. For the activation cross-sections of deuteron induced nuclear reactions on erbium only our experimental data sets exist, published up to 40 MeV (Hermanne et al., 2011b; Tarkanyi et al., 2016; Tárkányi et al., 2007b; Tárkányi et al., 2007c; Tárkányi et al., 2007a). In this work we extend the projectiles' energy range up to 50 MeV energy. The measured data are relevant for the general activation database and for further development of theoretical codes. Among the investigated products the $^{165}$Tm → $^{165}$Er, $^{169}$Er, $^{167}$Tm, $^{170}$Tm and $^{171}$Er → $^{171}$Tm radio nuclei are of interest in nuclear medicine.

# 2. Experimental

The excitation functions were measured by the activation method using stacked foil irradiation. The stack was irradiated at an external beam-line of the Cyclone 110 cyclotron of the Université Catholique in Louvain la Neuve (LLN) for 40 min with a 50 MeV, 50 nA deuteron beam. The stack contained a sequence of 10 blocks, each consisting of Al (10 μm), $Sr(NO_3)_2$ (3 μm, sedimented), Al (50 μm), Er (32 μm), Al (10 μm), $Ba(NO_2)$ (2 μm, sedimented), Al (50 μm) and Ag (10 μm) foils. The 10 Er targets covered the 50-32 MeV energy range (up to 40 MeV it was already published by our group). The targets were irradiated in a short Faraday cup. Gamma-ray spectra were measured with HPGe detectors coupled to Canberra multi-channel analyzers equipped with GENIE acquisition software. Four series of gamma-ray spectra were measured to follow the decay, started at different times after the end of bombardment, i.e. 8.9-10.7 h, 53.3-69.6 h, 268.1- 313.5 h and 1700.8- 1985.0 h, respectively. Gamma-ray spectra were evaluated by the automatic fitting algorithm included in the Genie 2000 package (Canberra, 2000)or in an iterative process using the Forgamma code (Székely, 1985). The decay data were adopted from NUDAT 2.6 (NuDat, 2014) and the Q values of the contributing reactions from the Q-value calculator (Pritychenko and Sonzogni, 2003) and are shown in Table 1.

The recommended data for simultaneously measured excitation functions of the $^{27}$Al(d,x)$^{22,24}$Na monitor reactions were taken from the IAEA database (Tárkányi et al., 2001)(last updated in 2007). As erbium is not monoisotopic, therefore so called elemental cross-sections were determined. Uncertainty of cross-sections was determined according to the recommendation (International-Bureau-of-Weights-and-Measures, 1993) by taking the sum in quadrature of all individual contributions: beam current (7%), beam-loss corrections (max. 1.5%), target thickness (1%), detector efficiency (5%), photo peak area determination and counting statistics (1-20 %).

The initial beam energies in targets (for stack preparation) were obtained from a degradation calculation based on the incident energy and the Andersen (Andersen and Ziegler, 1977) polynomial approximation of stopping powers. In order to fit with the recommended monitor reactions a correction was applied to the measured collected charge (final) (Tárkányi et al., 1991). Uncertainty of energy was estimated by following the cumulative effects of possible uncertainties (primary energy, target thickness, energy straggling, correction to monitor reaction) during the energy degradation.

Table 1. Decay characteristics of the investigated reaction products

| Nuclide | Half-life | $E_\gamma$ (keV) | $I_\gamma$ (%) | Contributing process | Q-value (MeV) |
|---|---|---|---|---|---|
| **$^{170}$Tm** | 128.6 d | **84.25** | 2.48 | $^{170}$Er(d,2n) | −3.32 |
| **$^{168}$Tm** | 93.1 d | **198.25** | 54.49 | $^{167}$Er(d,n) | 3.09 |
| | | 447.52 | 23.98 | $^{168}$Er(d,2n) | −4.69 |
| | | 631.71 | 9.26 | $^{170}$Er(d,4n) | −17.95 |
| | | 720.39 | 12.21 | | |
| | | 741.36 | 12.81 | | |
| | | 816.0 | 50.95 | | |
| | | 821.16 | 11.99 | | |
| **$^{167}$Tm** | 9.25 d | **207.8** | 42.0 | $^{166}$Er(d,n) | 2.68 |
| | | | | $^{167}$Er(d,2n) | −3.76 |
| | | | | $^{168}$Er(d,3n) | −11.53 |
| | | | | $^{170}$Er(d,5n) | −24.79 |
| **$^{166}$Tm** | 7.70 h | 80.59 | 11.5 | $^{166}$Er(d,2n) | −6.05 |
| | | 691.25 | 7.5 | $^{167}$Er(d,3n) | −12.48 |
| | | 705.33 | 11.1 | $^{168}$Er(d,4n) | −20.25 |
| | | **778.81** | 19.1 | $^{170}$Er(d,6n) | −33.52 |
| | | 785.90 | 10.0 | | |
| | | 1176.70 | 9.6 | | |
| **$^{165}$Tm** | 30.06 h | **242.92** | 35.5 | $^{164}$Er(d,n) | 2.05 |
| | | **297.68** | 12.7 | $^{166}$Er(d,3n) | −13.07 |
| | | | | $^{167}$Er(d,4n) | −19.51 |
| | | | | $^{168}$Er(d, 5n) | −27.28 |
| | | | | $^{170}$Er(d,7n) | −40.54 |
| **$^{163}$Tm** | 1.810 h | **104.32** | 18.6 | $^{162}$Er(d, n) | 1.46 |
| | | 241.31 | 10.9 | $^{164}$Er(d,3n) | −14.29 |
| | | | | $^{166}$Er(d,5n) | −29.42 |
| | | | | $^{167}$Er(d,6n) | −35.85 |
| | | | | $^{168}$Er(d,7n) | −43.62 |
| **$^{171}$Er** | 7.516 h | 111.62 | 20.5 | $^{170}$Er(d,p) | 3.46 |
| | | **308.29** | 64 | | |
| **$^{161}$Er** | 3.21 h | 211.15 | 12.2 | $^{162}$Er(d,p2n) | -11.43 |
| | | 314.77 | 2.49 | $^{161}$Tm decay | -15.51 |
| | | **826.6** | 64 | | |

Abundances (%): $^{162}$Er (0.14), $^{164}$Er (1.61), $^{166}$Er (33.6), $^{167}$Er (22.95), $^{168}$Er (26.8), $^{170}$Er (14.9)

Increase the Q-values if compound particles are emitted by: np-d, +2.2 MeV; 2np-t, +8.48 MeV; n2p-$^{3}$He, +7.72 MeV; 2n2p-$\alpha$, +28.30 MeV.

The energies in bold are used for evaluation.

# 3. Results and discussion

## 3.1 Excitation functions

The measured cross-sections are shown in Figs. 1-8 in comparison with the earlier experimental data and with the theoretical predictions. The numerical data are collected in Tables 2 - 3.

Table 2. Experimental cross-sections for the $^{nat}Er(d,x)^{170,168,167,166}Tm$ reaction

| | | $^{170}Tm$ | | $^{168}Tm$ | | $^{167}Tm$ | | $^{166}Tm$ | |
|---|---|---|---|---|---|---|---|---|---|
| **E** | **ΔE** | **σ** | **Δσ** | **σ** | **Δσ** | **σ** | **Δσ** | **σ** | **Δσ** |
| **MeV** | | **mb** | | | | | | | |
| 49.57 | 0.20 | 5.75 | 3.50 | 68.32 | 8.39 | 204.01 | 22.90 | 226.65 | 26.37 |
| 48.05 | 0.24 | 5.62 | 1.09 | 66.37 | 7.77 | 215.31 | 24.18 | 205.21 | 23.50 |
| 46.49 | 0.28 | 7.51 | 1.11 | 90.16 | 10.51 | 244.48 | 27.46 | 196.55 | 22.63 |
| 44.89 | 0.32 | 8.39 | 2.01 | 72.70 | 8.72 | 253.84 | 28.50 | 197.30 | 22.81 |
| 43.25 | 0.36 | 3.51 | 0.49 | 81.19 | 9.40 | 292.10 | 32.79 | 210.80 | 23.96 |
| 41.56 | 0.40 | 5.93 | 2.29 | 92.58 | 11.04 | 293.99 | 33.00 | 236.32 | 27.15 |
| 39.82 | 0.44 | 4.88 | 1.21 | 102.86 | 12.20 | 287.67 | 32.30 | 283.68 | 32.45 |
| 38.02 | 0.49 | 7.38 | 1.63 | 101.54 | 12.20 | 236.45 | 26.54 | 278.06 | 31.63 |
| 36.16 | 0.54 | 10.11 | 4.85 | 132.60 | 15.50 | 240.90 | 27.05 | 355.41 | 40.53 |
| 34.22 | 0.58 | 8.31 | 3.52 | 173.69 | 19.50 | 243.49 | 27.33 | 381.52 | 43.53 |
| 32.19 | 0.63 | 6.29 | 2.06 | 211.23 | 24.07 | 256.09 | 28.75 | 441.26 | 50.20 |

Table 3. Experimental cross-sections for the $^{nat}Er(d,x)^{165,163}Tm$, $^{171,161}Er$ reaction reactions

| | | $^{165}Tm$ | | $^{163}Tm$ | | $^{171}Er$ | | $^{161}Er$ | |
|---|---|---|---|---|---|---|---|---|---|
| **E** | **ΔE** | **σ** | **Δσ** | **σ** | **Δσ** | **σ** | **Δσ** | **σ** | **Δσ** |
| **MeV** | | **mb** | | | | | | | |
| 49.57 | 0.20 | 377.07 | 42.38 | 353.80 | 41.05 | 3.73 | 0.49 | 15.86 | 2.28 |
| 48.05 | 0.24 | 392.09 | 52.71 | 311.33 | 35.70 | 2.95 | 0.39 | 15.86 | 2.09 |
| 46.49 | 0.28 | 407.74 | 45.88 | 271.66 | 30.87 | 3.85 | 0.47 | 12.45 | 1.64 |
| 44.89 | 0.32 | 409.52 | 46.01 | 224.00 | 26.11 | 3.72 | 0.47 | 9.86 | 1.49 |
| 43.25 | 0.36 | 447.23 | 50.24 | 176.91 | 20.58 | 4.22 | 0.51 | 7.52 | 1.21 |
| 41.56 | 0.40 | 432.48 | 48.61 | 113.70 | 14.56 | 5.07 | 0.62 | 3.81 | 1.14 |
| 39.82 | 0.44 | 444.82 | 50.03 | 66.35 | 8.66 | 4.73 | 0.58 | 2.19 | 0.70 |
| 38.02 | 0.49 | 460.97 | 51.81 | 43.95 | 7.56 | 5.23 | 0.64 | | |
| 36.16 | 0.54 | 435.69 | 48.97 | 31.96 | 8.33 | 5.73 | 0.70 | | |
| 34.22 | 0.58 | 442.00 | 49.62 | 25.15 | 6.77 | 7.81 | 0.93 | | |
| 32.19 | 0.63 | 453.67 | 50.97 | 35.20 | 9.24 | 8.11 | 0.96 | | |

### 3.1.1 Production $^{171}$Tm ($T_{1/2}$ = 1.92 y)

In spite of our efforts, as in our previous works, it was impossible to deduce reliable cross-sections for the medically relevant $^{171}$Tm produced via the $^{170}$Er(d,n) reaction due to low cross sections and to the very low intensity of the low energy (66.7 keV) gamma-line.

### 3.1.2 Production of $^{170}$Tm ($T_{1/2}$ = 128.6 d)

$^{170}$Tm is produced in the $^{170}$Er(d,2n) reaction. The β-decay of $^{170}$Tm is followed by emission of a weak 84.3 keV gamma-line. The overlapping and contamination from the 84.94 keV $K_{\beta1}$ X-ray line from gamma-ray excitation of Pb (shielding of the detector) were subtracted, resulting in small residual intensities with large uncertainties. The resulting excitation function is shown in Fig. 1. The comparison with the TENDL-2015 on-line library shows good agreement.

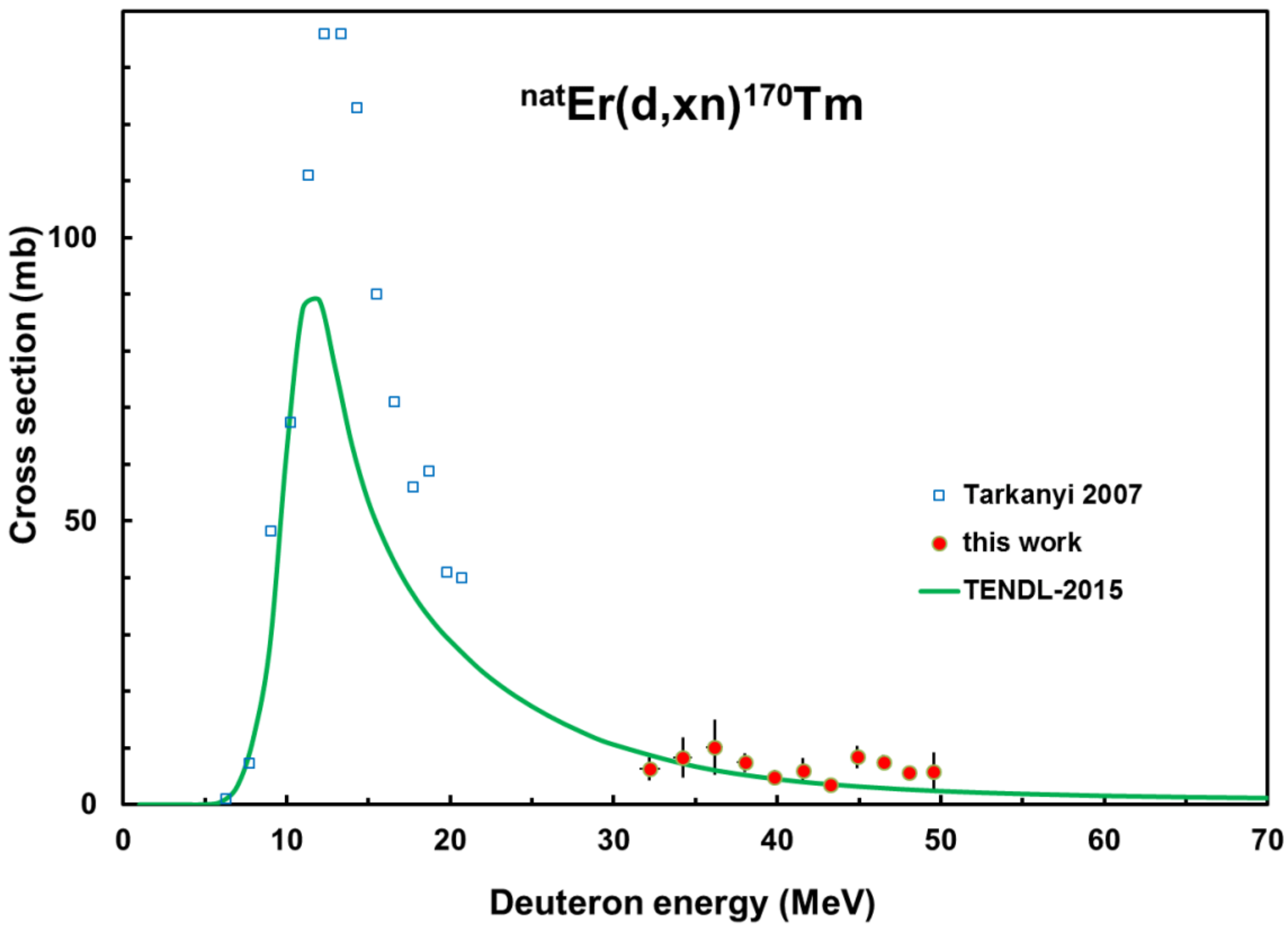

Fig. 1 Experimental and theoretical excitation functions of the $^{nat}$Er(d,xn)$^{170}$Tm reaction

### 3.1.3 *Production of $^{168}$Tm ($T_{1/2}$ = 93.1 d)*

The excitation function of $^{168}$Tm is shown in Fig. 2. The agreement with the earlier data is very good in the overlapping energy range. The theory describes well the experimental data, but at higher deuteron energies an obvious energy shift can be observed.

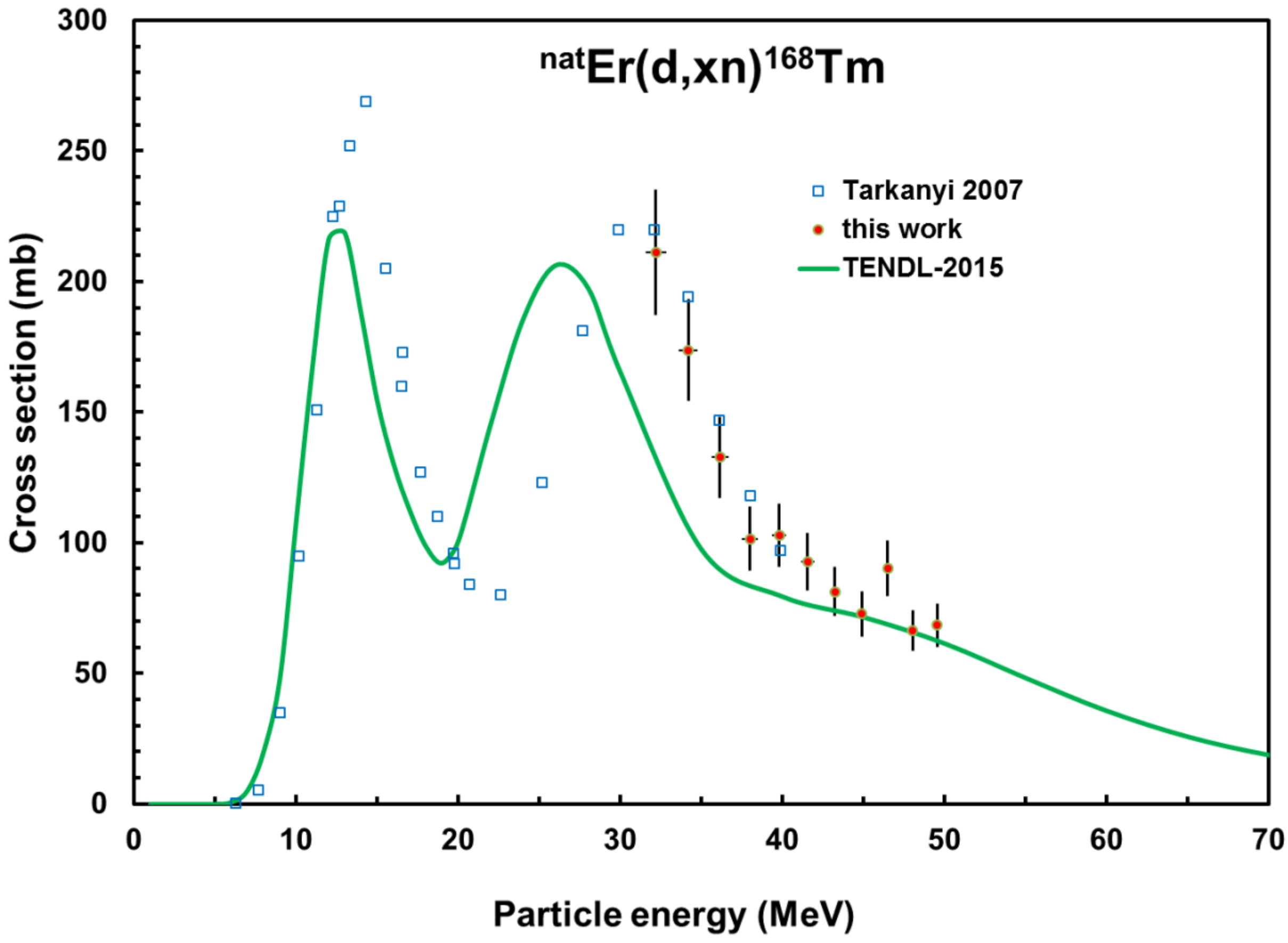

Fig. 2 Experimental and theoretical excitation functions of the $^{nat}$Er(d,xn)$^{168}$Tm reaction

### *3.1.4 Production of $^{167}$Tm ($T_{1/2}$= 9.25 d)*

The excitation function for production of $^{167}$Tm is shown in Fig. 3. The theory underestimates the experiment and again some obvious energy shift is observed at high energies.

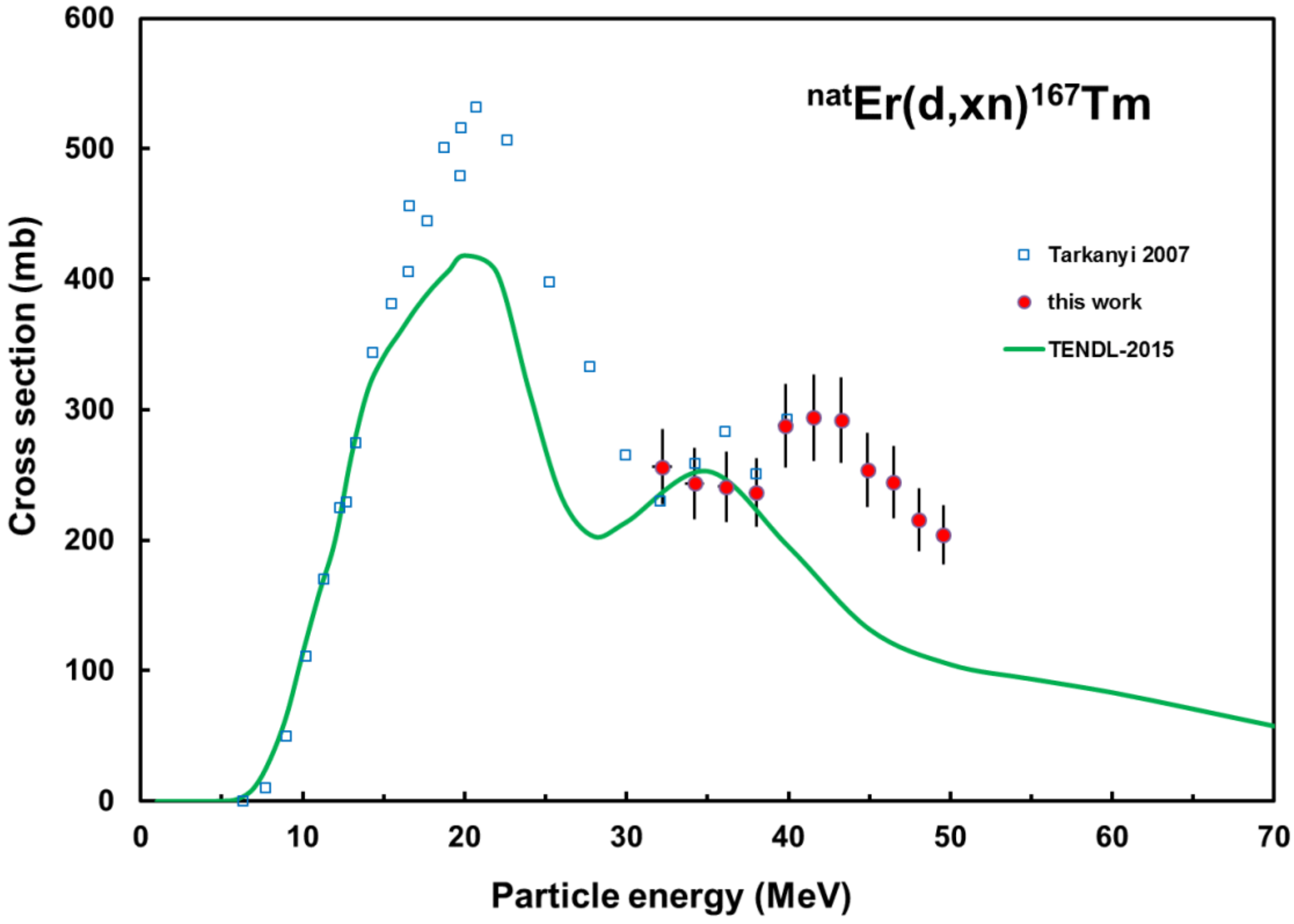

Fig. 3 Experimental and theoretical excitation functions of the $^{nat}Er(d,xn)^{167}Tm$ reaction

### *3.1.5 Production of $^{166}$Tm ($T_{1/2}$= 7.70 h)*

The cross-sections for production of the $^{166}$Tm radionuclide are shown on Fig 4. The new data are in acceptable agreement with our previous experimental data in the low energy range. The magnitude of the TENDL prediction is acceptable, but an energy shift is also indicated in this case.

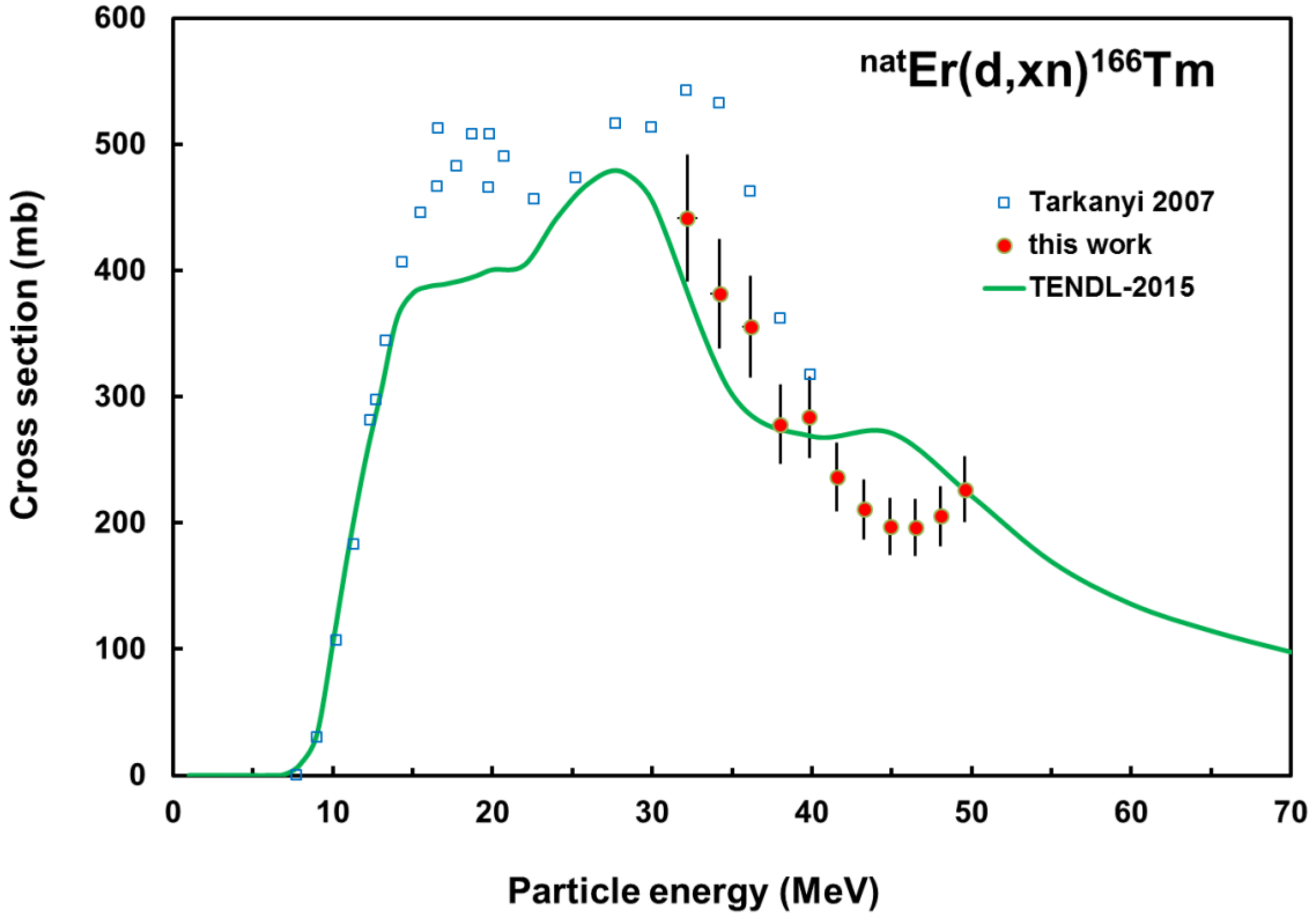

Fig. 4 Experimental and theoretical excitation functions of the $^{nat}$Er(d,xn)$^{166}$Tm reaction

### *3.1.6 Production of $^{165}$Tm ($T_{1/2}$= 30.06 h)*

According to Fig. 5 the new data are lower compared to our previous experimental data in the overlapping energy range, which may be caused by the relatively large uncertainties, but fit rather well with the TENDL-2015 calculations.

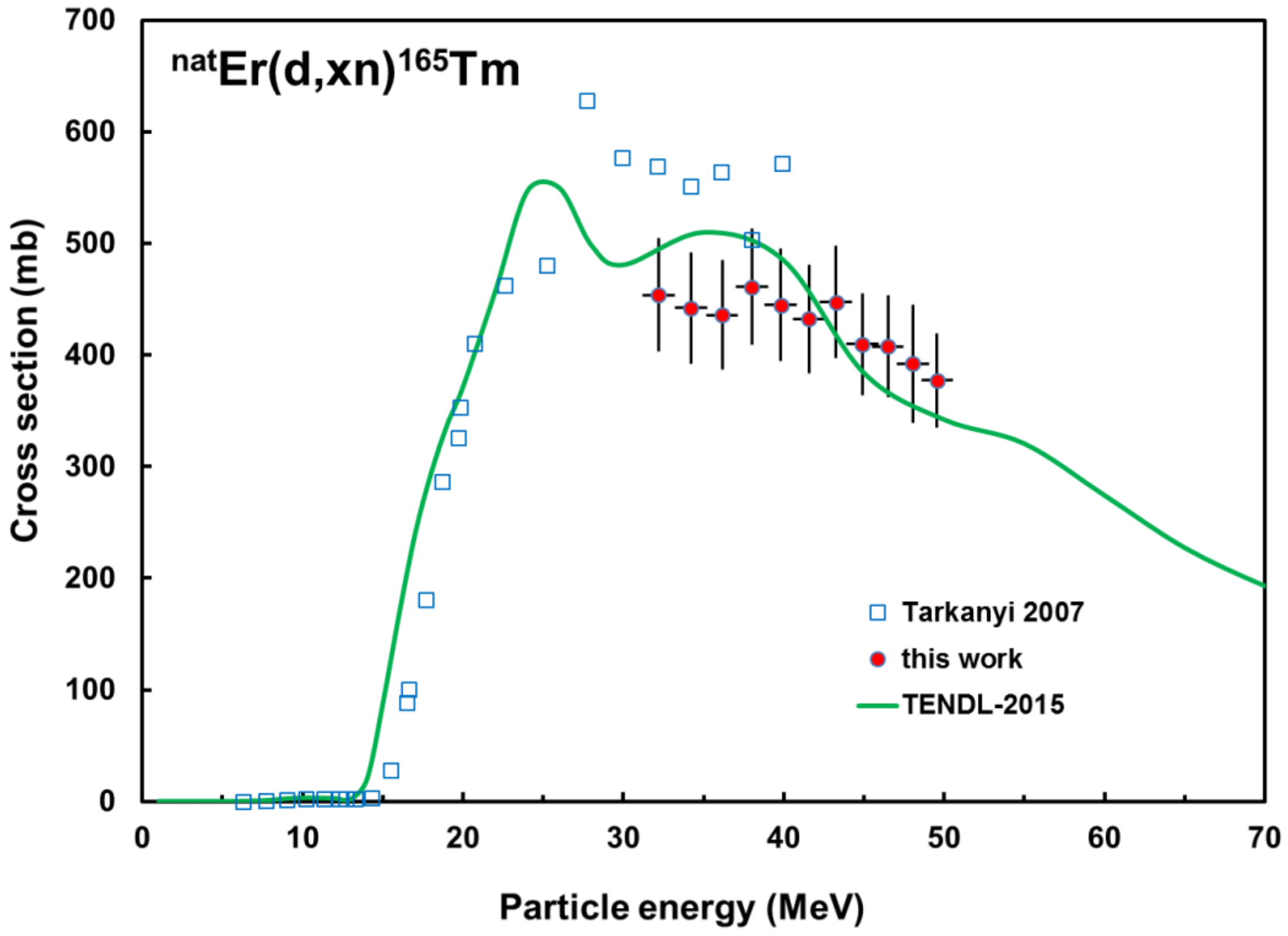

Fig. 5. Experimental and theoretical excitation functions of the $^{nat}$Er(d,xn)$^{165}$Tm reaction

### *3.1.7 Production of $^{163}$Tm ($T_{1/2}$ = 1.810 h)*

As obvious from the Fig. 6, a significant energy shift can be observed above 30 MeV between experiment and theory and above 50 MeV the experimental data become significantly higher.

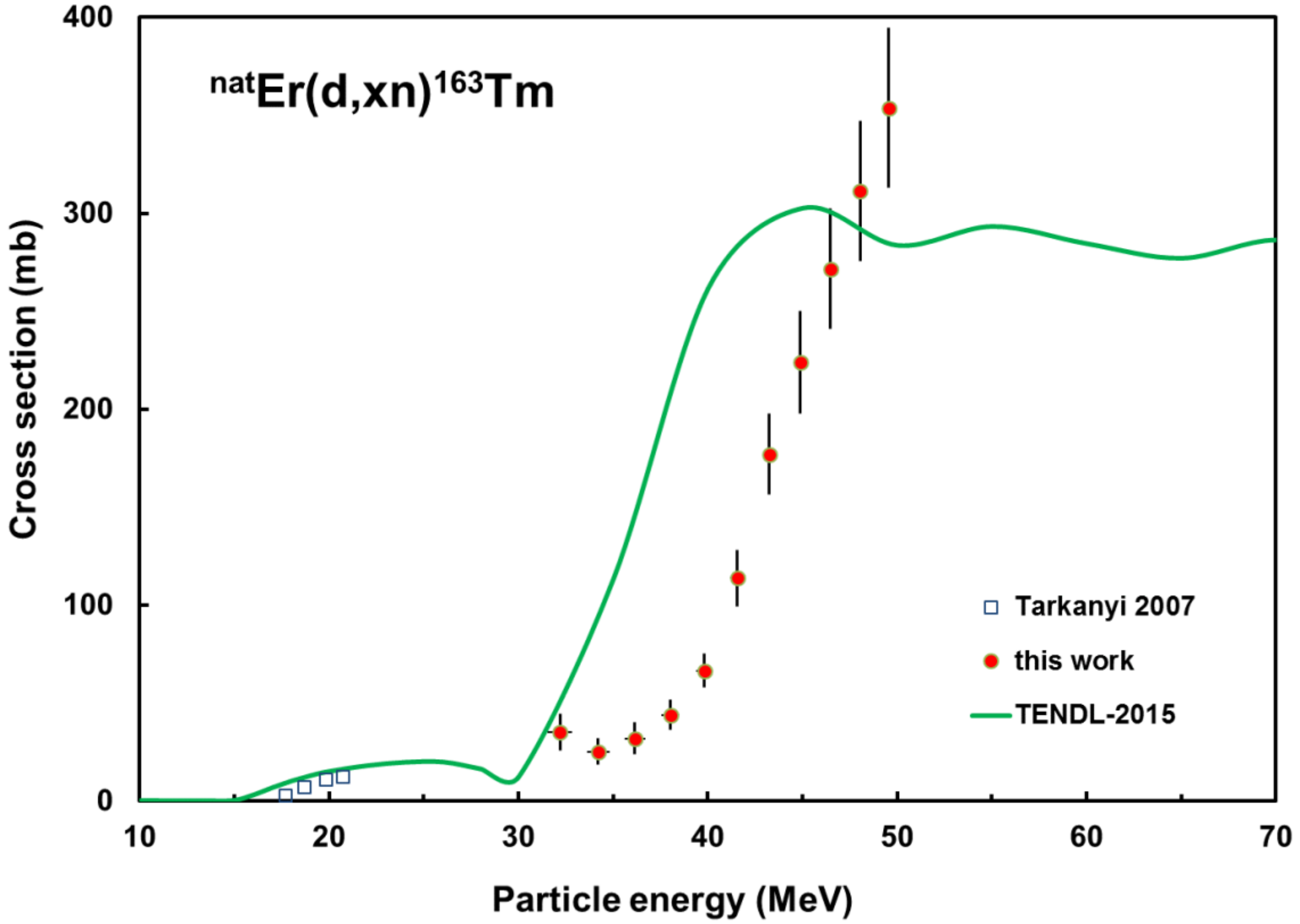

Fig.6 Experimental and theoretical excitation functions of the $^{nat}$Er(d,xn)$^{163}$Tm reaction

### *3.1.8 Production of $^{171}$Er ($T_{1/2}$ = 7.516 h)*

The prediction systematically underestimates both experimental data sets in the whole energy range (Fig. 7). The reason is the poor modeling of the deuteron breakup, which leads to underestimation of the $^{170}$Er(d,p) reaction by the used model (Khandaker et al., 2015; Khandaker et al., 2014), which is the only contributing reaction to the $^{171}$Er formation.

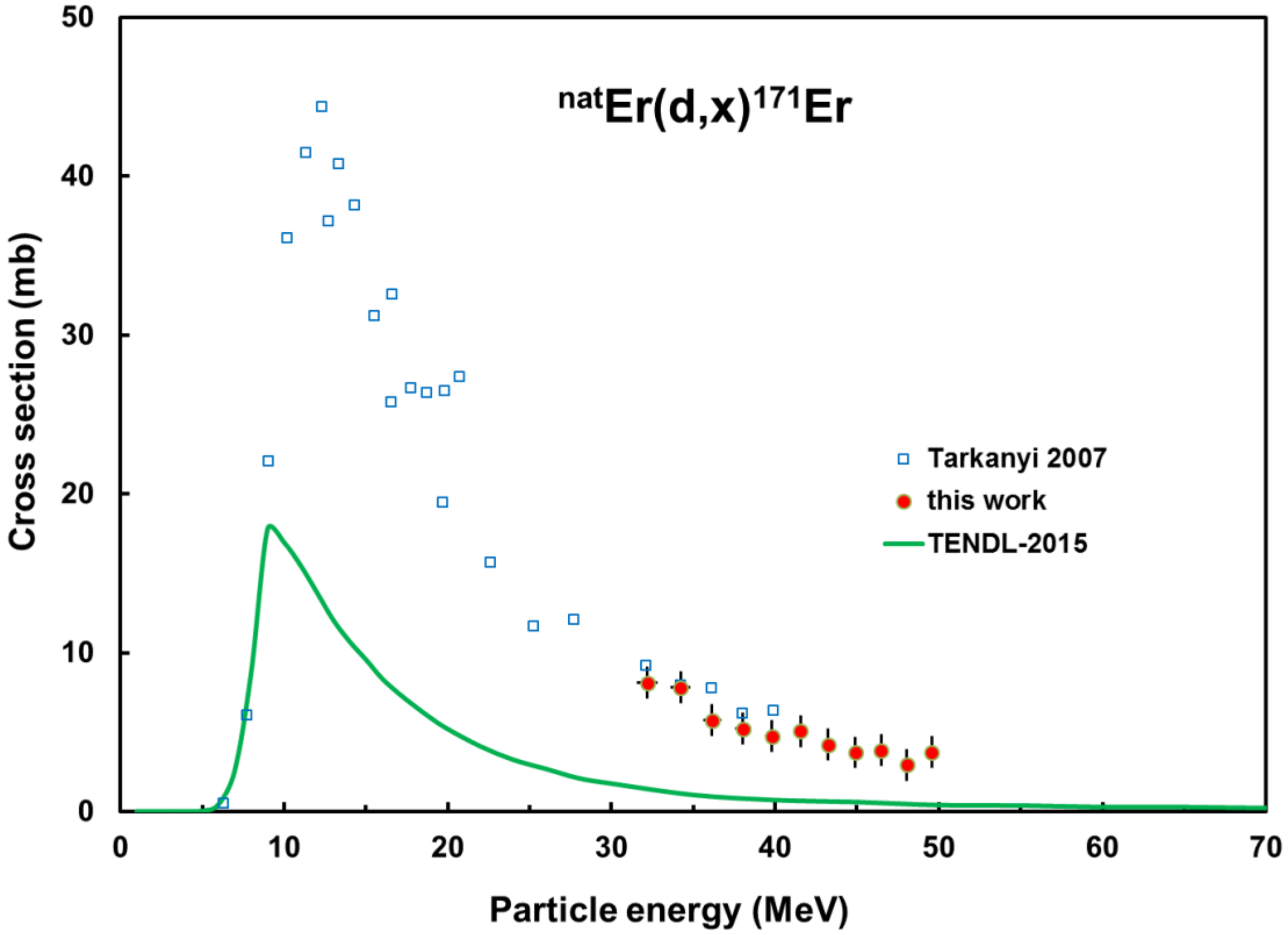

Fig. 7 Experimental and theoretical excitation functions of the $^{nat}$Er(d,x)$^{171}$Er reaction

### *3.1.9 Production of $^{161}$Er ($T_{1/2}$ = 3.21 h)*

For production of $^{161}$Er no earlier experimental cross section data were found. It is even not covered by our previous work because of the relatively low half-life and the low expected cross section in the previously measured energy region. The excitation function for cumulative production (from $^{161}$Tm, $T_{1/2}$= 30.2 min, ε=100%) is shown in Fig. 8 where the theory overestimates the experimental values up to 45 MeV, but a good agreement is seen above this value, which can also be the result of an energy shift.

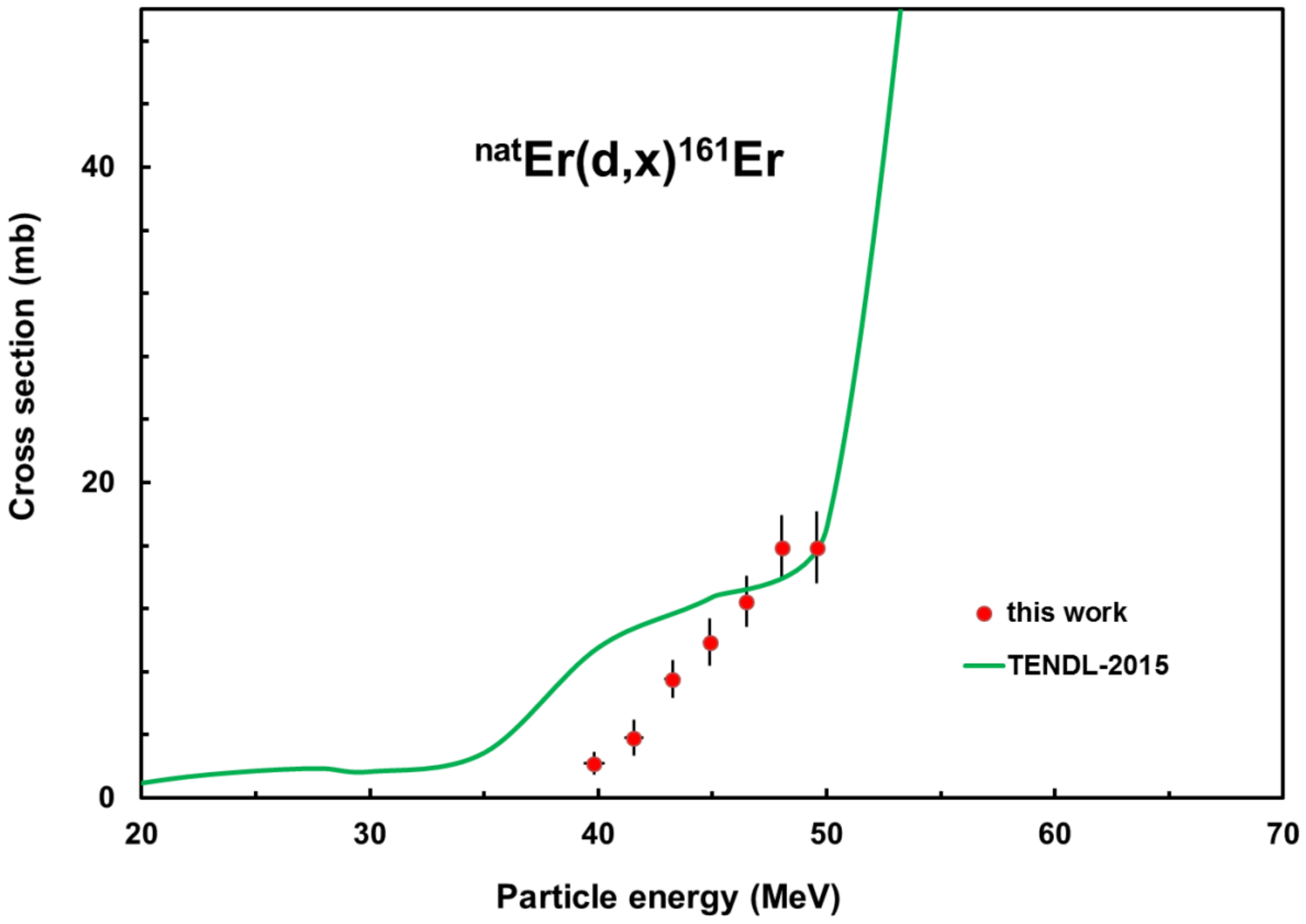

Fig. 8 Experimental and theoretical excitation functions of the $^{nat}$Er(d,x)$^{161}$Er reaction

## 4. Integral yields

Integral yields down to the reaction threshold as a function of the energy were calculated by using fitted curves to experimental cross-section data. The results for physical yields (Bonardi, 1987; Otuka and Takács, 2015) are presented in Fig. 9 in comparison with the experimental yield data in the literature (Dmitriev et al., 1982; Dmitriev et al., 1980). Comparing our new calculated results with the data from the literature the general conclusion is that the previous results of Dmitriev et al. are slightly lower than our new data except in the case of $^{163}$Tm, but their data are definitely higher at the lower energy region.

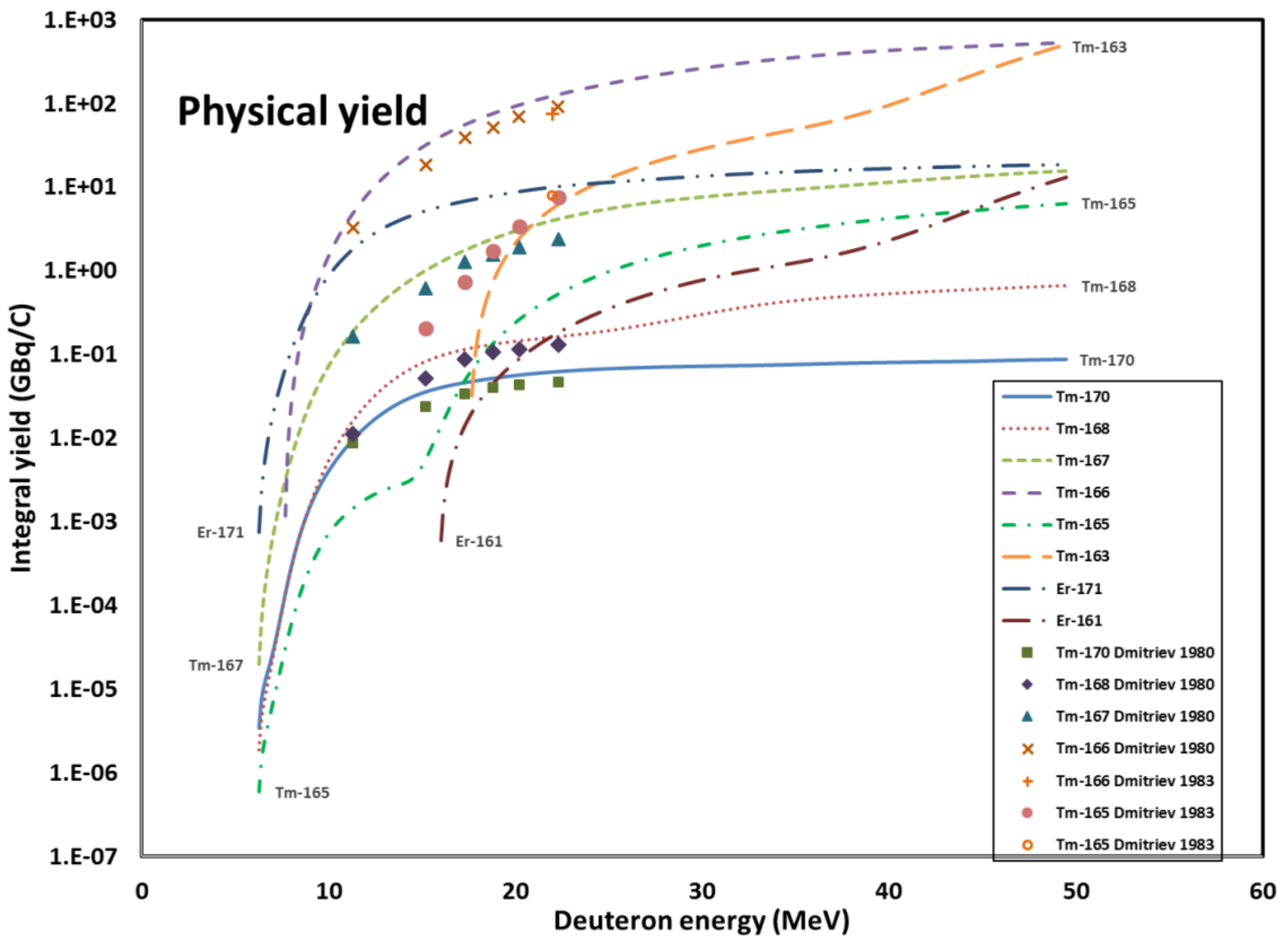

Fig. 9 Integral yields for production of the $^{163,165,166,167,168,170}$Tm and $^{171,161}$Er nuclear reactions deduced from the excitation functions in comparison with the literature experimental integral yields.

## 5. Summary and conclusions

Activation cross sections of the $^{nat}Er(d,x)^{163,165,166,167,168,170}Tm$ and $^{nat}Er(d,x)^{171,161}Er$ nuclear reactions have been measured experimentally in the 32-50 MeV energy range, above 40 MeV for the first time. The $^{nat}Er(d,x)^{161}Er$ reaction cross section data are reported here for the first time. The agreement with the earlier experimental data in the overlapping energy range is in most cases is acceptable, and with the theoretical predictions in the TENDL-2015 theoretical library only in several cases is acceptable. The discrepancy by the (d,p) reaction and the observed energy shift at higher energies suggest an improvement in the code. Comparison of charged particle production routes of medically relevant $^{167,170,171}$ Tm and $^{165,169}Er$ radioisotopes were discussed in detail in our previous works. By extending the energy range the present work gives important information on cross sections for production of medically relevant $^{161}Er$, $^{163}Tm$, $^{167}Tm$ and to test TENDL predictions in broader energy range.

### *Acknowledgement*

This work was done in the frame of MTA-FWO (Vlaanderen) research projects. The authors acknowledge the support of research projects and of their respective institutions in providing the materials and the facilities for this work.

## Figures

Fig. 1 Experimental and theoretical excitation functions of the $^{nat}Er(d,xn)^{170}Tm$ reaction

Fig. 2 Experimental and theoretical excitation functions of the $^{nat}Er(d,xn)^{168}Tm$ reaction

Fig. 3 Experimental and theoretical excitation functions of the $^{nat}Er(d,xn)^{167}Tm$ reaction

Fig. 4 Experimental and theoretical excitation functions of the $^{nat}Er(d,xn)^{166}Tm$ reaction

Fig. 5 Experimental and theoretical excitation functions of the $^{nat}Er(d,xn)^{165}Tm$ reaction

Fig.6 Experimental and theoretical excitation functions of the $^{nat}Er(d,xn)^{163}Tm$ reaction

Fig. 7 Experimental and theoretical excitation functions of the $^{nat}Er(d,x)^{171}Er$ reaction

Fig. 8 Experimental and theoretical excitation functions of the $^{nat}Er(d,x)^{161}Er$ reaction

Fig. 9 Integral yields for production of the $^{163,165,166,167,168,170}Tm$ and $^{171,161}Er$ nuclear reactions deduced from the excitation functions in comparison with the literature experimental integral yields.